\documentclass{aa}

\usepackage{graphicx}       
\usepackage{booktabs}       
\usepackage{amsmath}        
\usepackage{natbib}         
\usepackage{hyperref}       
\usepackage{txfonts}        
\usepackage{url}
\usepackage{xcolor}
\usepackage{soul}          

\newif\ifshowchanges
\showchangesfalse

\ifshowchanges
  \newcommand{\rev}[1]{{\color{blue}#1}}
  \newcommand{\del}[1]{{\color{red}\st{#1}}}
\else
  \newcommand{\rev}[1]{#1}
  \newcommand{\del}[1]{}
\fi

\begin{document}

\title{A measurement-based approach to EPFD calculation, to quantify the impact of satellite mega-constellations on low-frequency radio astronomy}

\titlerunning{Quantifying the impact of satellite mega-constellations on low frequency radio astronomy}


\author{
  Tingay, S.J.\inst{1}
  \and
  Hellbourg, G.\inst{2,3}
  \and
  Sheen, D.\inst{4,5}
  \and
  Grigg, D.\inst{1}
  \and
  Winkel, B.\inst{6}
  \and
  Di Vruno, F.\inst{7,8}
  \and
  Indermuehle, B\inst{9}
}

\institute{
  International Centre for Radio Astronomy Research, Curtin University, Bentley, WA, Australia;
  \email{s.tingay@curtin.edu.au}
  \and
  Cahill Center for Astronomy and Astrophysics, California Institute of Technology, Pasadena, CA 91125, USA
  \and
  Owens Valley Radio Observatory, California Institute of Technology, Big Pine, CA 93513, USA
  \and
  Department of Electrical Engineering and Computer Science, Massachusetts Institute of Technology, Cambridge, MA 02139, USA
  \and
  Haystack Observatory, Massachusetts Institute of Technology, Westford, MA 01886, USA
  \and
  Max-Planck-Institut f\"{u}r Radioastronomie, Radioobservatorium Effelsberg, Max-Planck-Strasse 28, D-53902 Bad Münstereifel, Germany
  \and
  IAU Centre for the Protection of the Dark and Quiet Sky from Satellite Constellation Interference, France
  \and
  SKA Observatory, Jodrell Bank, Macclesfield, SK11 9FT, United Kingdom
  \and
  CSIRO Space and Astronomy, PO Box 76, Epping, NSW 1710, Australia
}

\date{Received \today; accepted \today}

\abstract
  {
    Satellite mega-constellations have been shown to generate Unintended Electro-Magnetic Radiation (UEMR) that impacts low frequency radio astronomy.  The Radio Regulations (RR) of the International Telecommunications Union Radiocommunication Sector (ITU-R) contain the logical basis for addressing UEMR, but the enforcement mechanisms are absent.
  }
  {
    To adapt compatibility study methods based on satellite constellation simulations, in particular the Equivalent Power Flux-Density (EPFD) framework, to real-world radio telescope measurement data.
  }
  {
    We adapt the ITU-R EPFD calculation framework to radio astronomy measurement data obtained from all-sky interferometric imaging. We explicitly address the conceptual differences between forward-model EPFD calculations and measurement-based approaches, including antenna gain treatment, temporal averaging, and detection limitations.  In particular, we consider low frequency systems, which are typically wide field-of-view and interferometric instruments.
  }
  {
      For the first time, we demonstrate that because all-sky interferometric observations can be interpreted as a set of simultaneous beamformed measurements they enable a direct adaptation of the EPFD formalism to measurement data. Applying this framework to low-frequency observations, we confirm previous estimations \citep{2023A&A...676A..75D} that the resulting measurement-based EPFD distributions exceed the radio astronomy interference threshold levels defined in Recommendation ITU-R RA.769-2 in approximately 50--70\% of samples, depending on polarization and observing subset. These exceedance fractions are well above the 2\% single-system data-loss criterion in Recommendation ITU-R RA.1513-2. Expressed \del{as a} \rev{using the conventional} 2\% compatibility margin, \del{the 98th percentile of the distributions are 16.3--18.6~dB above the RA.769 threshold.} \rev{the corresponding margins are negative, ranging from $-18.6$ to $-16.3$~dB, equivalent to a required uniform attenuation of 16.3--18.6~dB.}
  }
  {
    We show that measurement-based implementations provide a practical pathway toward verification and long-term monitoring at low radio frequencies, which could form an important component of enforceability.  We suggest instrumentation approaches to support measurement. 
  }


\maketitle

\section{Introduction}
\label{sec:intro}


The recent growth of mega-constellations of Earth-orbiting satellites for the purposes of global communications is having well documented impacts on both ground-based and space-based astronomical measurements \citep{adhikary2026impact,hellbourg2025quiet,borlaff2025satellite}.  Further, regulatory filings that propose an even greater rate of growth indicate that the impacts to astronomy are likely to increase in the future.

One particular impact on astronomy that has emerged is the presence of so-called Unintended Electro-Magnetic Radiation (UEMR), first detected at low radio frequencies (in the VHF band) from the Starlink constellation of satellites, emanating from onboard electronics and electric sub-systems, i.e., not related to purposeful transmissions \citep{2023A&A...676A..75D,2023A&A...678L...6G,2024A&A...689L..10B,2024arXiv241214483G,2025A&A...698A.244Z,2025A&A...699A.307G}. While Starlink is not expected to be particularly unique in this respect as a constellation, it has been the major first-mover in the industry and has, therefore, the most obvious impacts at the current point in time.



The term UEMR was first coined by \cite{2023A&A...676A..75D}. The terminology remains under active discussion within the ITU-R community, where alternative terms such as Unintentional Radio Frequency Energy (URFE) have recently been discussed. In this paper, we retain the term UEMR for consistency with the existing astronomical literature. Within the ITU-R framework, the Radio Regulations primarily refer to wanted and unwanted emissions. Wanted emissions are the signals within a necessary frequency bandwidth transmitted via an antenna of any application under a recognized radio service, while unwanted emissions are the inevitable by-products (such spectral sidelobes, harmonics, or intermodulation products) of the wanted-emission operations. More generally, the term `radiation' is used for any electromagnetic wave, including but not limited to emissions. Thus, it also applies to topics of Electromagnetic Compatibility (EMC; see in particular Articles 15.12 and 15.13 of the RR).

In this work, we use the term UEMR to refer specifically to the broader term `radiation' in the ITU-R terminology, while excluding emissions, consistent with previous astronomical literature. For example, radiation generated by power converters or clocks/oscillators not associated with generation of the carrier signal that are part of the suite of electronics on the satellite platform would be considered UEMR, as opposed to downlink transmissions via a directional antenna.

One of the challenges for radio astronomy is to present analyses of the impacts observed in the framework, language, and terminology of the regulatory world, including that of the ITU.  Differences in nomenclature, differences in approaches to modeling and measurement, and differences in instrumentation can represent a barrier to communicating impact and reaching a mutual appreciation between astronomers, regulators, and satellite operators.

The aim of this paper is to contribute to the closure of some of these gaps, in order to start to adapt astronomical measurements and measurement systems such that they can be utilised and understood in an ITU framework.

In particular, we consider the calculation of Equivalent Power Flux Density (EPFD) within the ITU-R framework as applied to mega-constellations, adapted as a tool for evaluating impacts on low frequency radio astronomy when considering UEMR specifically.  We consider the adaptation of this calculation, designed around forward modeling simulations, to measurement data and describe the inevitable mismatches between model-driven calculations and measurement-driven, instrument-specific realisations of those calculations. This motivates us to propose a measurement-based implementation of the EPFD framework, suitable for the analysis of real-world radio astronomy observations and complementary to the model-driven EPFD calculations currently described within the ITU-R.

An accepted and standardised measurement-based EPFD calculation would be a valuable verification tool for astronomers, satellite operators, and regulators in the future, in order to systematically monitor the future impacts of UEMR from mega-constellations as the situation inevitably evolves quickly, due to: the establishment of more and larger constellations; the introduction of mitigation measures; and the establishment of any future regulations.  In short, systematic monitoring will be required to assess impacts and attribute any positive or negative changes to the future actions of astronomers, operators, or regulators.  The shared responsibility to meet the needs of all stakeholders should start from a solid evidence base.  To be clear, we do not advocate that such a measurement-driven calculation replace the current EPFD methodology which, in many cases, is the only tool available for understanding the impacts due to proposed mega-constellations, but as a complementary approach that has advantages specifically when treating UEMR. In addition, it has to be noted, that the in-situ measurement of the EPFD over the whole visible sky at any radio telescope is today only practical at low frequencies with synthetic aperture arrays.

Thus, this paper proceeds with a description of the ITU-R EPFD calculation in Section \ref{sec:epfd-calc}, outlining the forward modeling approach currently utilised to predict the impacts of satellite constellations on radio astronomy.  In Section \ref{EPFD-inter}, we interpret all-sky interferometric imaging in an EPFD context.  In Section \ref{sec:epfd-meas} we then describe the limitations of this approach if one wishes to consider real-world measurement information.  In Section \ref{sec:epfd-ill} we adapt the ITU-R EPFD calculation in application to measurements and illustrate this by example, utilising an already published extensive set of UEMR measurements of Starlink satellites in the VHF band.  We focus on measurements in ITU protected bands for radio astronomy.  We describe both the data and results, comparing results to the previous literature.  Motivated by these results, in Section \ref{sec:prop} we propose a measurement-driven EPFD calculation that could complement the theoretical calculation employed so far, potentially using it for verification purposes.  Moreover, we suggest particular realisations of radio astronomy instrumentation that could efficiently and effectively feed such a calculation with data, resulting in a measurement system and methodology that could support ongoing systematic monitoring and verification to inform future actions.  Finally, in Section \ref{sec:disc} we discuss our overall results and outline the main conclusions of our work.

\section{The Equivalent Power Flux-Density (EPFD) calculation for radio astronomy as defined by the ITU-R}
\label{sec:epfd-calc}
\subsection{ITU-R Recommendations and their role in EPFD studies}

The evaluation of interference to the Radio Astronomy Service (RAS) within the ITU-R framework relies on a set of complementary Recommendations that define both the relevant thresholds and the methodology used to assess compliance. Among these, Recommendations ITU-R RA.769 \citep{ITU_RA769} and RA.1513 \citep{ITU_RA1513} provide the fundamental criteria for determining detrimental interference, while Recommendations ITU-R S.1586 \citep{ITU_S1586} and M.1583 \citep{ITU_M1583} describe how EPFD calculations are performed for different satellite services.

Recommendation ITU-R RA.769 defines the threshold levels of power flux density that correspond to detrimental interference in radio astronomy observations. \rev{In this work, we use the continuum thresholds from RA.769, rather than the spectral-line thresholds.} These limits are provided on a per-band basis and are expressed in a range of units (in this paper, we use the limits expressed in dB(W\,m$^{-2}$\,Hz$^{-1}$)), assuming a 0\,dBi receiving antenna and an integration time of 2\,000\,s as typical parameters in ITU-R studies). These thresholds form the reference against which EPFD calculations are ultimately compared.

Recommendation ITU-R RA.1513 complements this by defining acceptable levels of data loss in radio astronomy observations. In particular, it specifies that, for a given satellite system, the RA.769 threshold should not be exceeded for more than a small fraction of integration periods, and that the aggregate contribution of all systems should remain below the threshold for at least 95\,\% of EPFD samples (and below 98\,\% of epfd samples for a single system e.g. Starlink). The interpretation of the EPFD calculation is therefore in terms of cumulative distribution functions, where compliance is assessed by the fraction of samples exceeding the threshold.

The computation of EPFD itself is described in Recommendations ITU-R S.1586 and M.1583, which apply to different classes of satellite services. Recommendation ITU-R S.1586 primarily addresses fixed-satellite service (FSS) systems, while Recommendation ITU-R M.1583 applies to mobile-satellite and radionavigation-satellite services. Although the Recommendations differ in their intended applications and compliance methodologies, they share a common computational framework. In both cases, the sky is discretized into cells, the interference contribution from all satellites is evaluated for each cell as a function of time, and the resulting EPFD values are averaged over typical radio astronomical integration times (i.e. 2000~s). The calculations are repeated many times in order to account for the statistical variability associated with satellite motion and observational geometry, thereby producing cumulative distribution functions suitable for compatibility assessment.

Despite the apparent complexity of these documents, the core concept is relatively simple: EPFD studies aim to quantify how often the interference produced by satellite systems exceeds the sensitivity limits of radio astronomy, as defined by RA.769. The comparison is inherently statistical, reflecting both the time variability of satellite constellations and the directional dependence of the interference. The methodology described in this paper follows the same principles, but replaces simulated satellite contributions with measured data, providing an empirical perspective on the EPFD framework.

\subsection{Definition of EPFD}

Equivalent Power Flux-Density (EPFD) is a synthetic quantity used to describe the aggregate interference from multiple transmitters moving across the sky. It is defined as the power flux density that, if received from a hypothetical point source located at the boresight of the radio astronomy antenna, would produce the same total received power as the sum of all interfering signals. In this sense, EPFD provides a convenient way to collapse a complex, distributed interference environment into a single, direction-independent metric.

\subsection{Core mathematical formulation}


For $N_a$ visible non-geostationary satellites, the instantaneous EPFD in the sky cell $n$ is given by
\begin{equation}
\mathrm{EPFD}(\boldsymbol{\omega}_n) = \sum_{i=1}^{N_a}
    \frac{P_i \, G_{t,i}(\boldsymbol{\theta}'_i, \boldsymbol{\varphi_i}) \, G_r(\boldsymbol{\theta}_i, \boldsymbol{\omega}_n)}
         {4\pi d_i^2 \, G_{r,\max}},
\label{eq:epfd}
\end{equation}

\noindent where $P_i$ is the radiated power of satellite $i$ in the reference bandwidth, $G_{t,i}(\boldsymbol{\theta}'_i, \boldsymbol{\varphi_i})$ is the transmit antenna gain of the i\textsuperscript{th} satellite in direction $\boldsymbol{\theta}'_i$  (towards the receive antenna or radio astronomy telescope) when the satellite antenna (or its main beam) points to $\boldsymbol{\varphi}_i$, and $G_r(\boldsymbol{\theta}_i, \boldsymbol{\omega}_n)$ is the radio telescope gain pattern in the direction of the satellite $\boldsymbol{\theta}_i$ while pointing to sky cell $n$ at $\boldsymbol{\omega}_n$. All angles except $\boldsymbol{\omega}_n$ depend on time as satellites are in motion. $G_{r,\max}$ is the maximum gain of the receive antenna, and $d_i$ is the slant range between the transmit antenna and the receive antenna.

Because radio astronomy protection criteria are referenced to a 0\,dBi receive gain, this expression is typically reformulated without the normalization by $G_{r,\max}$ as
\begin{equation}
\mathrm{EPFD}_{0\,\mathrm{dBi}}(\boldsymbol{\omega}_n) = \sum_{i=1}^{N_a}
    \frac{P_i \, G_{t,i}(\boldsymbol{\theta}'_i, \boldsymbol{\varphi_i}) \, G_r(\boldsymbol{\theta}_i, \boldsymbol{\omega}_n)}
         {4\pi d_i^2}.
\label{eq:epfd0}
\end{equation}

\noindent Unlike EPFD, $\mathrm{EPFD}_{0\,\mathrm{dBi}}$ is referenced to a hypothetical receiving antenna with 0~dBi gain. This allows direct comparison with the threshold levels defined in Recommendation ITU-R RA.769, which are similarly expressed relative to a 0~dBi receiving antenna. The quantity remains a power flux-density metric at the Earth’s surface and does not itself include the wavelength-dependent effective aperture of a receiving antenna. This formulation is common to both Recommendations ITU-R S.1586 and M.1583.

At higher frequencies (e.g. above a few GHz) the equations would need to be modified to incorporate additional propagation effects, such as atmospheric attenuation. In this case, the equation becomes
\begin{equation}
\mathrm{EPFD}_{0\,\mathrm{dBi}}(\boldsymbol{\omega}_n) = \sum_{i=1}^{N_a}
    \frac{P_i \, G_{t,i}(\boldsymbol{\theta}'_i, \boldsymbol{\varphi_i}) \, G_r(\boldsymbol{\theta}_i, \boldsymbol{\omega}_n)}
         {4\pi d_i^2} T_{\mathrm{atm}}(\boldsymbol{\theta}_i),
\label{eq:epfd0_atm}
\end{equation}

\noindent where $T_\mathrm{atm}(\boldsymbol{\theta}_i)$ is the transmission coefficient through the atmosphere as a function of the apparent direction to satellite i from the radio telescope $\boldsymbol{\theta}_i$, and may be calculated from an atmospheric model such as that given in Recommendation ITU-R P.676-13. Additional path effects such as diffraction by terrain at the horizon may be incorporated similarly, however, only the path loss due to distance from the satellite is explicitly considered in the current ITU framework.


\subsection{Inputs to EPFD simulations}

EPFD calculations are typically performed as forward-model simulations of satellite constellations. They require detailed knowledge of the system under study, including the orbital configuration and visibility of satellites, their wanted or  unwanted emission characteristics and transmit antenna patterns, and the pointing behavior when applicable. In particular, modern broadband communication systems such as SpaceX/Starlink feature a large number of synthesized beams which are dynamically steered to cells on the ground using complex resource allocation schemes. Both the antenna patterns and the pointing schemes are usually considered proprietary, and are therefore generally unavailable to simulations. On the receiving side, the telescope location and antenna response pattern must be specified, usually through high-resolution side-lobe models. Propagation effects are generally limited to free-space loss for space-to-Earth links at low frequencies. Because satellite interference is highly time-variable, the calculation is inherently dynamic and is evaluated over a standard interval of 2\,000\,s.

\subsection{From instantaneous EPFD to compliance metrics}

The transformation from instantaneous EPFD values to regulatory compliance metrics follows a well-defined statistical procedure:

\begin{enumerate}
    \item The visible sky is discretized into cells of approximately equal solid angle (typically $\sim$9\,deg$^2$);
    
    \item For each cell, the telescope is assumed to point to a random direction in that cell, and the instantaneous $\mathrm{EPFD}_{0\,\mathrm{dBi}}$ is computed as a function of time (e.g. at 1\,s resolution) over a 2\,000\,s interval;
    
    \item For each cell, a single EPFD sample is obtained by averaging the instantaneous values over the 2\,000\,s window;
    
    \item The process is repeated over many trials, varying the initial time and relative geometry between the constellation and the telescope pointing within each cell;
    
    \item All resulting samples are combined to form an empirical cumulative distribution function (CDF), restricted to physically observable directions (e.g. above a minimum elevation); and
    
    \item Compliance is assessed by comparing the fraction of samples exceeding the RA.769 threshold with the limits defined in Recommendation ITU-R RA.1513.
\end{enumerate}

\rev{Physically, each EPFD sample represents a 2\,000~s averaged interference estimate for a particular telescope pointing direction and constellation geometry. The exceedance fraction therefore represents the fraction of such pointing/time samples for which the interference level is above the RA.769 detrimental-interference threshold. Within the RA.1513 framework, this fraction is interpreted as a proxy for the percentage of radio astronomy data loss caused by detrimental interference.}

This procedure, described in detail in Annex~3 of Recommendation ITU-R S.1586 and Annex~2 of Recommendation ITU-R M.1583, forms the basis of EPFD-based compatibility assessments for satellite systems.

In addition to reporting the fraction of samples exceeding the RA.769 threshold, it is useful to express the result as a margin relative to the 2\% single-system criterion in Recommendation ITU-R RA.1513. We define this margin as
\rev{\begin{equation}
\Delta_{2\%} = T_{\mathrm{RA.769}} - Q_{98}\!\left(\mathrm{EPFD}_{0\,\mathrm{dBi,dB}}\right),
\label{eq:margin_2pct}
\end{equation}}
where $Q_{98}$ is the 98th percentile of the EPFD distribution and $T_{\mathrm{RA.769}}$ is the relevant RA.769 threshold. \del{A positive value of $M_{2\%}$ corresponds to the uniform reduction, in dB, that would be required for the EPFD distribution to satisfy the 2\% criterion.} \rev{With this convention, a positive margin indicates that the 98th percentile lies below the threshold, while a negative margin indicates that attenuation is required. The corresponding required uniform attenuation is $A_{2\%}=-\Delta_{2\%}$ when $\Delta_{2\%}<0$.} This interpretation is valid for a uniform reduction of all source flux densities, since the EPFD calculation is linear in source power.

\section{Application to all-sky interferometric observations}
\label{EPFD-inter}

\subsection{Measurement-based interpretation of EPFD using interferometric imaging}

The application of the EPFD formalism to measurement data requires a directional interpretation of the received signal. In conventional radio astronomy observations at higher frequencies (>300 MHz), telescopes (generally dishes) are intrinsically directional, with narrow primary beams. As a result, a measurement based implementation of the EPFD using a single dish telescope would require directional measurements (equivalent to the cells defined in the calculation) for the defined period of time, covering all the cells in the sky. As a result, observational scanning strategies would be required, leading to prohibitively long measurement campaigns.

In contrast, low-frequency interferometric arrays operate with much larger fields of view. This is a direct consequence of the inverse relationship between wavelength and antenna size; at low frequencies, the physical size of individual antenna elements is small compared to the wavelength, resulting in broad primary beams. When combined in an interferometric array, these elements enable the reconstruction of images over a large fraction of the visible sky, often approaching full-sky coverage in a single snapshot observation.

A key property of interferometric imaging is that each pixel in the reconstructed image corresponds to the response of a synthesized beam formed in that direction \citep{thompson2017interferometry}. In other words, an all-sky image can be interpreted as the result of beamforming simultaneously in all directions. This provides a natural mapping to the EPFD formalism, which is defined for a directional receiving antenna.

More precisely, the value of a given pixel does not represent emissions originating exclusively from that direction, but rather the output of a beamformed response; it contains contributions from the entire sky, weighted by the synthesized beam pattern. This pattern has a maximum gain in the direction of interest and attenuates signals from other directions, although this attenuation is not infinite. As a result, each pixel can be interpreted as the response of a directional antenna pointed toward that direction, with finite sidelobes contributing residual signals from elsewhere in the sky. 
The value of each pixel therefore contains contributions from receiver thermal noise, diffuse sky emission, and interfering sources across the visible sky, weighted by the synthesized beam response and modulated by the primary beam of the individual antenna element.

While this interpretation suggests that each image pixel represents an instantaneous measurement of the aggregate interference weighted by the synthesized beam response, practical limitations remain. In the idealized case of perfect calibration and negligible thermal and sky noise, the pixel value directly contains the combined contribution of all sources entering through the receiving response of the array. In practice, however, receiver noise, diffuse sky emission, calibration uncertainties, and unresolved background structure introduce a fluctuating noise floor that limits the sensitivity of the method, particularly for weak contributions entering through distant sidelobes.

Consequently, satellite signals can be identified through differential measurements relative to the surrounding background, either spatially across neighboring pixels or temporally as the source moves through the field. The methodology therefore measures the detectable excess power associated with moving emitters above the local background environment.

The key advantage provided by all-sky imaging is that signal detections can be obtained simultaneously across the full visible sky. This enables the construction of an instantaneous sky distribution of detectable interfering emissions, subject to the sensitivity limits of the telescope and the adopted detection threshold. Rather than estimating the EPFD contribution of individual satellites from reconstructed transmitter properties, the measured source distribution is weighted by the direction-dependent receiving response of the array. The resulting measurement-based EPFD maps therefore represent the aggregate contribution of all detected emitters, including their redistribution through the synthesized-beam sidelobes and the primary-beam response.


An additional consideration is the relationship between the synthesized beamwidth and the discretization of the sky into cells. The sky image pixels used in the source measurement should have an angular separation matched to the effective beamwidth of the synthesized beam (e.g. the sky should be fully sampled but not oversampled). Denser sky sampling will result in pixels that are not independent, which makes local background subtraction less sensitive and effectively reduces the source SNR, as the same signal strongly contributes to multiple pixels. Conversely, with sparse sampling, the peak beam response to a signal will be reduced on average. This sampling need not directly correspond to the cell size with which the EPFD is computed, however, EPFD cells substantially larger than a beamwidth may result in an estimated distribution that is artificially smoothed. Ensuring an appropriate match between beamwidth and cell size is therefore recommended for maintaining consistency with the assumptions underlying the EPFD framework, although the number statistics will likely not be affected significantly unless beam and cell sizes are significantly different.

Conceptually, the methodology follows the same directional summation principle as the conventional ITU-R EPFD calculation, but replaces simulated satellite contributions with measured sky brightness distributions derived from interferometric observations. Rather than reconstructing individual satellite transmitter properties, the measured sky distribution is directly weighted by the direction-dependent receiving response of the telescope in order to estimate the aggregate EPFD distribution.

To obtain time-averaged EPFD values consistent with ITU-R definitions, the detected flux densities are first accumulated in sky cells and averaged over 2\,000\,s windows, yielding an average source flux density $\bar{\Phi}_i$ for each source cell $i$. These measured flux densities already include the combined effects of satellite transmitter power, propagation loss, and transmit antenna response. The remaining EPFD operation is therefore the receive-side directional weighting:
\begin{equation}
\widehat{\mathrm{EPFD}}_{0\,\mathrm{dBi}}(\boldsymbol{\omega}_n)
=
\sum_{i=1}^{N_{\mathrm{cells}}}
\bar{\Phi}_i
G_r(\boldsymbol{\theta}_i,\boldsymbol{\omega}_n),
\label{eq:epfd0_meas}
\end{equation}
where $\boldsymbol{\theta}_i$ is the direction of source cell $i$, $\boldsymbol{\omega}_n$ is the output pointing direction, and $G_r(\boldsymbol{\theta}_i,\boldsymbol{\omega}_n)$ is the receive gain toward the source direction when the array is phased toward the output cell. In the present implementation, this response is modeled as the product of the primary-beam gain in the source direction, the normalized synthesized-beam response between the source and output directions, and the coherent array gain. Repeating this procedure over independent 2\,000\,s averaging windows yields the EPFD distributions used for compatibility assessment.

By leveraging the equivalence between interferometric imaging and beamforming in all directions, it is possible to bridge the gap between theoretical EPFD formulations and real-world observations.

\subsection{Interpretation as a measurement-based EPFD estimator}

The methodology described above can be interpreted formally as a measurement-based estimator of the EPFD quantity defined within the ITU-R framework. In the conventional forward-model approach, the EPFD is obtained by summing the contributions from all satellites after weighting by the receiving antenna response. In the present work, the same operation is instead applied to a measured sky brightness distribution derived from interferometric imaging.

The estimator may therefore be written as
\begin{equation}
\widehat{\mathrm{EPFD}}_{0\,\mathrm{dBi}}(\boldsymbol{\omega}_n)
=
\sum_{i=1}^{N_{\mathrm{cells}}}
\bar{\Phi}_i
G_r(\boldsymbol{\theta}_i,\boldsymbol{\omega}_n),
\end{equation}
where $\bar{\Phi}_i$ is the measured average flux density associated with sky cell $i$, and $G_r(\boldsymbol{\theta}_i,\boldsymbol{\omega}_n)$ is the direction-dependent receiving response between source cell $i$ and output direction $\boldsymbol{\omega}_n$. In the array implementation used here, $G_r(\boldsymbol{\theta}_i,\boldsymbol{\omega}_n)$ is evaluated in the source direction $\boldsymbol{\theta}_i$ for an array phased toward $\boldsymbol{\omega}_n$. The primary-beam factor is therefore associated with the incident source direction, while the synthesized response describes the pointing-dependent array response between $\boldsymbol{\omega}_n$ and $\boldsymbol{\theta}_i$.

In the idealized case of perfect calibration, complete source detection, exact beam knowledge, negligible thermal noise, and negligible temporal smearing, the expectation value of the estimator converges toward the true EPFD:
\begin{equation}
\mathbb{E}
\left[
\widehat{\mathrm{EPFD}}_{0\,\mathrm{dBi}}(\boldsymbol{\omega}_n)
\right]
=
\mathrm{EPFD}_{0\,\mathrm{dBi}}(\boldsymbol{\omega}_n).
\end{equation}
In this limit, the methodology provides an unbiased empirical estimate of the EPFD distribution.

In practice, the estimator is affected by a number of observational selection effects and instrumental limitations that modify the recovered EPFD distributions relative to the true interference environment. The dominant effects affecting the methodology are summarized in Table~\ref{tab:selection_effects}, and elaborated upon in Section \ref{sec:epfd-meas}. Importantly, most of these effects act preferentially in the direction of reducing the recovered aggregate interference power. As a consequence, the resulting EPFD distributions should generally be interpreted as conservative lower-bound estimates of the true interference environment rather than as overestimates.

\begin{table*}[t]
\centering
\begin{tabular}{p{3.3cm} p{4.7cm} p{2.2cm} p{5.5cm}}
\hline
\textbf{Effect} &
\textbf{Origin} &
\textbf{Bias sign} &
\textbf{Impact on EPFD estimate} \\
\hline

Detection threshold &
Finite instrument sensitivity and source detection limits &
Negative &
Weak emissions below the detection threshold are omitted from the analysis, reducing the recovered aggregate interference power. \\

Temporal smearing &
Finite image integration time combined with satellite motion &
Negative &
Motion-induced smearing reduces the apparent peak brightness of moving emitters, lowering detection completeness and recovered source power. \\

Elevation cut &
Restriction to elevations above 20$^\circ$\del{ (or 40$^\circ$ in the restricted analysis)} &
Negative &
\del{Low-elevation c}\rev{C}ontributions \rev{below 20$^\circ$ elevation} are excluded from the estimator, removing signals entering through low-gain regions of the receiving response. \\

Incomplete satellite association &
Unmatched or rejected detections &
Negative &
Some valid satellite emissions may remain unidentified or excluded from the analysis. \\

Finite observing cadence &
Observations restricted to discrete observing periods and frequency settings &
Negative &
Emissions occurring outside the sampled observing intervals or frequency bands are not included in the estimator. \\

Beam model uncertainty &
Imperfect knowledge of the synthesized or primary beam response &
Positive or negative &
Can locally overestimate or underestimate the directional weighting applied during the EPFD estimation. \\

Flux-density calibration uncertainty &
Systematic calibration errors in measured source fluxes &
Positive or negative &
Directly rescales the recovered EPFD distributions. \\

\hline
\end{tabular}
\caption{Principal selection effects affecting the measurement-based EPFD estimator.}
\label{tab:selection_effects}
\end{table*}

At the same time, the methodology provides a direct empirical constraint on the aggregate interference actually experienced by a radio telescope, naturally incorporating real propagation effects, antenna responses, observational geometry, and temporal variability in a manner that is difficult to reproduce accurately in purely theoretical forward-model calculations.

\section{Limitations on the verification of the EPFD calculation via measurement}
\label{sec:epfd-meas}

While the EPFD framework provides a well-defined method for simulating and quantifying interference from satellite systems, its verification using measurement data is subject to important limitations arising from differences between theoretical assumptions and observational reality, in particular when considering UEMR.

An important practical difference between model-driven EPFD calculations and measurement-based implementations arises from the finite integration time of interferometric imaging. In the ITU-R framework, the position and contribution of each satellite are assumed to be known exactly at every instant, and the resulting EPFD samples are subsequently averaged over intervals such as 2\,000\,s to produce the required cumulative distributions. In contrast, radio astronomy images are themselves formed from visibilities integrated over finite time intervals. During these integrations, satellites move across the sky, causing their signals to decorrelate and smear spatially in the reconstructed images. This smearing reduces the apparent peak flux density of moving emitters by distributing their power over multiple pixels and sky cells, thereby reducing detection completeness, particularly for weak sources near the sensitivity limit.

Astronomical sky emission is normally protected from such smearing through phase tracking, which compensates for the known sidereal motion of the sky. Extending this approach to satellite signals is substantially more complex, since each satellite follows an independent trajectory that may not be known precisely a priori. Accurate compensation would therefore require independent phase tracking for many simultaneous moving sources on uncertain tracks, significantly increasing processing complexity and data rates. Consequently, satellite signals remain partially smeared in practical all-sky imaging observations, and this effect must be considered as an intrinsic sensitivity limitation of the measurement-based methodology.

Measurements are affected by both instrument noise and the astronomical sky.
Signal of interest detection therefore requires thresholds, typically defined in terms of signal-to-noise ratio, introducing selection effects and non-uniform sensitivity due to time-variable and spatially dependent noise. Techniques such as differencing consecutive images can mitigate contributions from the instrument noise floor and the astronomical background. However, the sky is not strictly stationary, and residuals from imperfect differncing remain. In addition, the astronomical background itself is structured and can be bright, particularly near the Galactic plane, making the separation of satellite signals incomplete. As a result, detection completeness is reduced, especially for weak emissions, and the measured source distribution contains only signals above the detection threshold while excluding weak sources, effectively yielding a lower bound on the true interference environment.

Another limitation arises from the identification of signals. The EPFD forward calculation assumes perfect knowledge of satellite positions, while measurements require the association of detections with specific satellites using orbital parameters. Imperfections in orbit determination, timing, or fitting can lead to missed or incorrect associations, resulting in an incomplete accounting of the contributions from a given constellation.

Polarization introduces further complexity. Measurements are generally made in two orthogonal polarizations, while the EPFD formalism does not explicitly address polarization. Each polarization can be treated as an independent measurement, but combining them assumes consistent source properties, which may not hold for UEMR. It is therefore preferable to analyse polarization channels separately.

Despite these limitations, measurement-based approaches offer a key advantage; they encapsulate several poorly known quantities, such as transmitter power and gain pattern, into a single observable, the received flux density. This is particularly important for UEMR, where emission mechanisms and transmit characteristics are not well understood and cannot be reliably modeled.  The measurement-based EPFD approach is therefore most useful precisely where the forward-model EPFD calculation is least applicable, because for UEMR there is no $G_{t}(\theta_{i})$ or $P_{i}$ to utilise in the calculation.

Overall, the methodology provides a direct empirical estimate of EPFD subject to observational selection effects and instrument-specific limitations, which must be considered when interpreting the resulting distributions.

\section{Illustrative application of the EPFD calculation to measurement data from large-scale, all-sky low frequency surveys}
\label{sec:epfd-ill}

The analysis script used in this work is openly available online\footnote{\url{https://github.com/greghell/EPFD/blob/main/epft_CDF.py}}.

\subsection{The data}
\label{sec:eda2-data}


The set of measurements we utilise for this analysis has already been published and are the result of a large-scale blind survey of UEMR
by \citet{2025A&A...699A.307G}.  We briefly describe the data, but refer the reader to the cited paper, and references therein, for further details.  It is important to note here that this survey was not designed or conducted with the analysis contained in this paper in mind.  The analysis here is therefore opportunistic and, as will be clear later in the paper, any future measurement campaign should be specifically designed to support the analysis presented below.

All-sky images with a cadence of 2 s were collected using the Engineering Development Array (Version 2: EDA2, \cite{10.1117/1.JATIS.8.1.011010}) at 24 different centre frequencies over the course of 29 days of observing, over a frequency range corresponding to that used by the low frequency Square Kilometre Array (SKA).  At each centre frequency, approximately 0.9 MHz of bandwidth was used to create the images, for both of the native linear polarisations of the EDA2 antennas (XX = east-west oriented dipole and YY = north-south oriented dipole).  With 256 dual-polarised antennas on a circular footprint of approximately 35 m diameter, the EDA2 produces images with a zenith angular resolution of several degrees at a frequency of 150 MHz.  The total survey dataset includes approximately 76 million all-sky images, across all frequencies, observations, and polarisations.  The dataset is a rich and unbiased record of signals above the horizon from the EDA2 location at the SKA site in Western Australia.  The data are fully publicly available at the following link: \url{https://zenodo.org/records/15089853}.

For the analysis below, we focus on measurements made at frequencies that overlap with frequency bands protected for radio astronomy by the ITU, listed in Table \ref{tab:tab1}, and the 150.05-153~MHz band more specifically. The detections analyzed in this band are most naturally interpreted as unintended radiation from satellite platform electronics. There are no allocated satellite transmissions in this frequency range or adjacent that could give rise to these signals through out-of-band
emissions. As such, the measurements considered here probe a class of emissions that is inherently difficult to model within the conventional EPFD framework.

\begin{table*}
\centering
\begin{tabular}{lcccc}
\hline
\textbf{Frequency band (MHz)} & \textbf{Region 1} & \textbf{Region 2} & \textbf{Region 3} & \textbf{Relevant footnotes} \\
\hline

13.36 -- 13.41 & P & P & P & 5.149 \\
25.55 -- 25.67 & P & P & P & 5.149 \\
37.5 -- 38.25 & S & S & S & 5.149 \\
73.0 -- 74.6 & --- & P & --- & 5.149 \\

150.05 -- 153.0 & P & --- & --- & 5.225 : P (Australia, India), 5.149 \\

322 -- 328.6 & P & P & P & 5.149 \\
406.1 -- 410.0 &P & P & P & 5.149 \\

608 -- 614 & \--- & P & --- & $\dagger$\\

\hline
\end{tabular}
\\$\dagger$ 5.304: P (African Broadcasting Area, ABA), 5.305: P (China), 5.306: S (Region 1 except ABA, Region 3 except China and India, 5.307: P (India)

\caption{
Radio Astronomy Service (RAS) allocations below 1~GHz in the ITU Radio Regulations.
Regions refer to the ITU geographical divisions: Region~1 (Europe, Africa, Middle East), Region~2 (the Americas), and Region~3 (Asia-Pacific).
Entries indicate primary (P) or secondary (S) allocations in the Table of Frequency Allocations (Article~5).
Country or area-specific allocations introduced via footnotes (e.g. 5.225) apply only to the listed administrations and do not constitute global allocations.
Some of the bands listed under footnote~5.149 are not formally allocated to RAS but are subject to a recommendation that administrations take all practicable steps to protect the service.
}
\label{tab:tab1}
\end{table*}

As noted in Section \ref{sec:epfd-meas}, the measurements we utilise are subject to real-world effects, compared to the forward-modeling approach described in the ITU-R for the EPFD calculation.  Measurements are conducted over a period in time (2 seconds in this case) and are not quasi-instantaneous as stipulated in the EPFD calculation.  The effect is that our measurements are tagged for a single time and location on the sky (the average over the measurement period) but in fact the signals may occupy more than the one cell stipulated in the EPFD calculation framework. Furthermore, very strong incident power levels could saturate the receiving system and thus lead to non-linearities in the amplifiers (creating spectral features such as harmonics or intermodulation products) or cause cross-talk in the beam-former \citep[see][]{2023A&A...676A..75D}.

Measurements are based on a method of detection and subject to detection thresholds. In this case, the differencing of pairs of closely spaced images is used to remove static (by generally good approximation) astronomical emission and isolate the signals of interest from satellites in motion.  Detections are made in the difference images against a defined signal-to-noise threshold.  This truncates the distribution of signal strengths accessible for use in the calculation to the top end of the signal strength distribution.  Moreover, the differencing methodology and the presence of non-static sources of astronomical emission (the Sun and occasionally ionospheric-induced scintillation of strong extragalactic radio sources) introduces systematic contributions to the image ``noise'', in addition to the thermal noise contribution.  The practical effect of this is that the signal-to-noise thresholds can vary with time of day, observing conditions, and source coordinates, truncating the signal strength distribution in a non-uniform and time variable manner.

The differencing approach starts to fail for signals at low elevations, where satellites have a low angular speed for the observer and the differencing timescale required to isolate the signal results in too much angular motion of astronomical signals.  This leads to an elevation limit of 20$^{\circ}$ being imposed on the survey. This elevation cut removes a significant fraction of the visible sky and therefore biases the measurement-based EPFD statistics toward higher elevations, where detections are more readily accessible.

Additionally, in order to undertake a calculation relevant to a particular mega-constellation, detection in itself is not enough.  Detections need to be positively identified with a unique satellite member of a particular constellation (in this case, the survey only considers the Starlink constellation).  Again, thresholds apply in the identification step (an angular matching radius, for example).  The survey utilised predictions based on publically available orbital information (TLEs from \url{space-track.org}) in the identification process.  Thus, if TLE information was out of date or inaccurate (for example, due to a maneuver or orbit change), detections sometimes could not be converted to identifications and therefore could not be used in a calculation.  This real-world limitation removes signals-of-interest from the distribution in a way that is not correlated with signal strength, such that even the strongest signals present in the data may not be usable in a measurement-driven EPFD calculation.

All of these real-world limitations on measurements need to be recognised and considered when undertaking an analog of the EPFD calculation with measurement data.  However, in all cases, such effects result in a reduced  calculation output, as the real-world effects represent the loss of signals from the calculation. The resultant EPFD distribution estimate is thus strictly a lower limit on the true distribution.



\subsection{Methods}
\label{subsec:meth}



\begin{figure*}[t]
    \centering
    \includegraphics[width=0.49\textwidth]{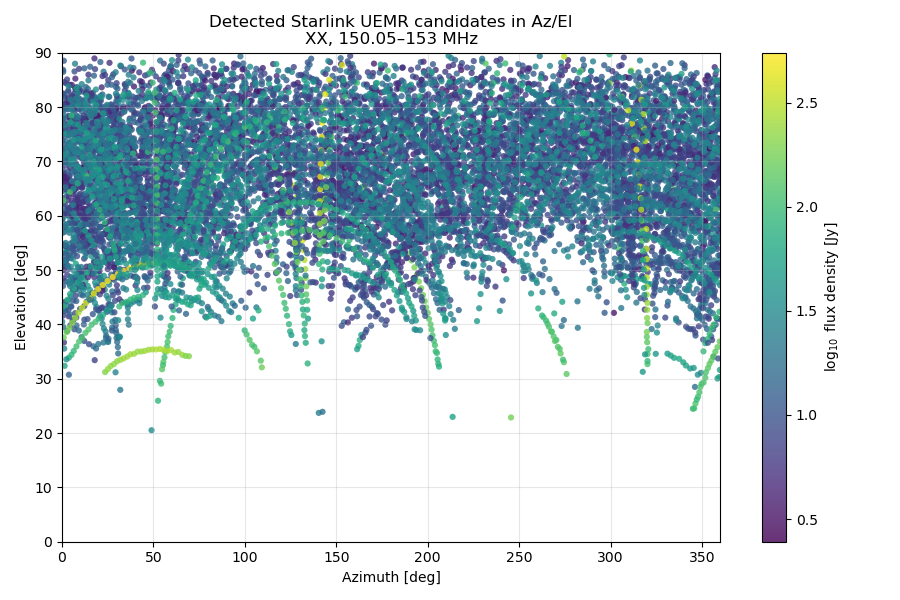}
    \includegraphics[width=0.49\textwidth]{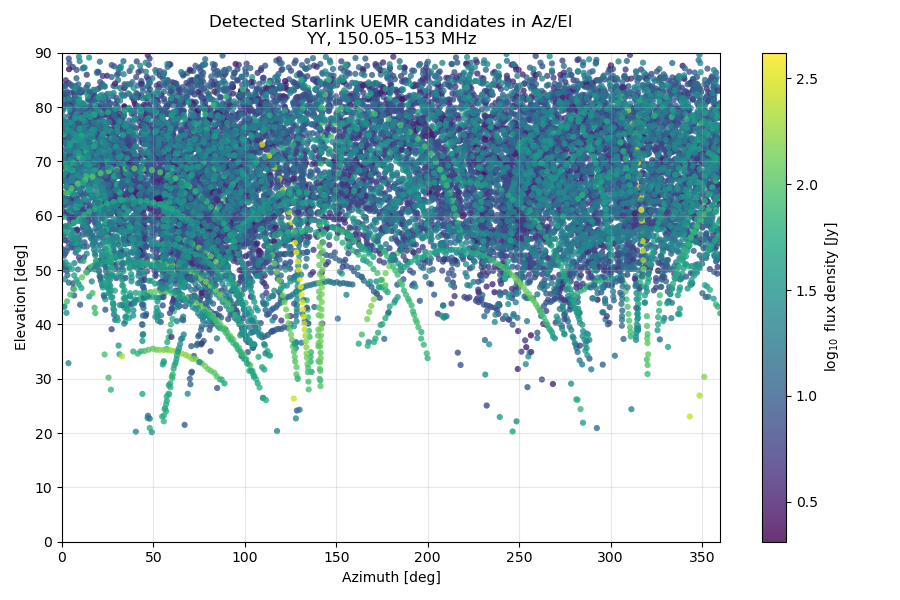}
    \caption{
    Distribution of detected and identified Starlink UEMR candidates in azimuth and elevation for the XX (left) and YY (right) polarizations. The detections are drawn from observing channels overlapping the 150.05--153~MHz protected band, including one channel that partially extends above the upper band edge.
    The color scale indicates the measured flux density in Jy. 
    The non-uniform distribution reflects both the orbital geometry of the constellation and observational selection effects associated with sensitivity and beam response.
    }
    \label{fig:azel_flux_distribution}
\end{figure*}

The detection catalog is filtered to retain measurements relevant to the analysis. This includes selecting full-band detections, applying an elevation cut, and optionally filtering by solar elevation to distinguish day and night conditions. For the frequency selection, we retain observing channels overlapping the 150.05--153~MHz protected band. In practice, this leaves detections at centre frequencies of 150.7812 and 153.125~MHz. The former lies fully within the protected band, while the latter partially overlaps the upper band edge, spanning approximately 152.68--153.58~MHz. We therefore interpret the results as representative of the protected-band environment near 150~MHz, rather than as a strictly in-band-only estimate. The two instrumental polarizations (XX and YY) are treated independently throughout the analysis, effectively forming two parallel realizations of the measurement process. The spatial distribution of detections and their measured flux densities are shown in Figure~\ref{fig:azel_flux_distribution} for both instrumental polarizations.

Each detected signal is assigned to a sky cell based on its fitted azimuth and elevation. The catalogue flux densities are used as source-extracted, beam-corrected estimates of the incident source flux densities, and no additional primary-beam correction is applied at this stage.

For each 2\,s integration, the contributions of all detections falling within the same sky cell are summed, producing instantaneous cell-based measurements of the detectable interference sky distribution. These samples are grouped into 2\,000\,s windows (1\,000 $\times$ 2\,s), consistent with the EPFD averaging interval. Within each window, the contributions in each cell are averaged over time, producing an average flux density, $\bar{\Phi}_i$, associated with each sky cell over the averaging interval.

These averaged cell fluxes do not yet represent the measurement-based EPFD distributions themselves. Rather, they constitute the measured source distribution to which the direction-dependent receiving response of the telescope is applied. Through this redistribution step, the contribution of each detected source is mapped across all output sky cells according to the synthesized-beam response and the primary-beam response in the source direction, thereby accounting for contributions entering through sidelobes as well as through the main beam.
Time steps within an active window that do not contain detections are retained as zero-valued samples in the averaging process. As a result, values are averaged over the full window duration rather than over the number of detections, ensuring that intermittent detections within a window are properly diluted in the time-averaged estimate.

Equation~\ref{eq:epfd0_meas} represents the measurement-based analogue of the standard EPFD summation. In the conventional ITU-R framework, the contribution of each satellite is weighted by the receiving antenna gain toward the considered sky direction and summed with the contributions from all other satellites. In the present work, the same operation is applied to the measured sky distribution obtained from the interferometric images. We emphasize that the primary-beam factor is evaluated in the source direction, not the output pointing direction. This is consistent with the EPFD definition, in which the receive gain is the gain of the telescope toward the interfering satellite for the pointing direction under consideration.

For each sky cell containing a measured average flux density, $\bar{\Phi}_i$, the contribution to every output sky direction is weighted by the direction-dependent response of the array. This response is modeled as the product of the synthesized beam and the primary beam response of the antenna elements. The resulting measurement-based EPFD maps therefore represent the aggregate contribution of all detected emitters after propagation through the receiving response of the telescope, including both main-beam and sidelobe contributions.

The receiving response used in the measurement-based EPFD summation is direction-dependent, since the synthesized beam of an interferometric array varies as a function of sky direction. For each signal-containing sky cell, the array response toward all output sky directions is computed using the synthesized-beam response between the source and output directions, combined with the primary-beam response of the antenna elements evaluated in the source direction.

This redistribution step is distinct from the image formation process itself. The interferometric images provide source-extracted flux-density estimates for the detected emitters, while the measurement-based EPFD summation accounts for the aggregate response of the telescope to those sources through both the main beam and synthesized sidelobes. In this sense, the redistribution applies the same receive-side directional weighting used in the EPFD formalism, but with measured source flux densities as inputs.

More formally, the normalized synthesis beam is computed as the squared magnitude of the array factor, scaled to unity at its peak:
\begin{equation}
\hat{B}_{\mathrm{synth}}(\mathbf{\hat{c}}\,|\,\mathbf{\hat{s}})
=
\frac{1}{N^2}
\left|
\sum_{n=1}^{N}
\exp\!\left[
j\,k\,\mathbf{r}_n\cdot
\left(\mathbf{\hat{c}}-\mathbf{\hat{s}}\right)
\right]
\right|^2,
\end{equation}
where $j=\sqrt{-1}$, $\mathbf{r}_n$ denotes the position of antenna $n$, $N$ is the number of array elements, $k$ is the wave number, and $\mathbf{\hat{s}}$ and $\mathbf{\hat{c}}$ are unit vectors corresponding to the source and output directions, respectively.



With this normalization, the receive response used in the EPFD summation may be written as:

\begin{equation}
G_r(\mathbf{\hat{s}},\mathbf{\hat{c}})
=
N\,\hat{B}_{\mathrm{synth}}(\mathbf{\hat{c}}\,|\,\mathbf{\hat{s}})
\, B_{\mathrm{prim}}(\mathbf{\hat{s}}),
\end{equation}

where $\hat{B}_{\mathrm{synth}}$ is the peak-normalized synthesized-beam response, $B_{\mathrm{prim}}$ is the primary-beam gain of the antenna elements evaluated in the source direction, and $N$ is the coherent array gain.
The factor $1/N^2$ in $\hat{B}_{\mathrm{synth}}$ normalizes the synthesized-beam power pattern to unit peak response; the absolute coherent array gain applied in the receive response is therefore $N$.

Substituting this expression into Eq.~\ref{eq:epfd0_meas} gives
\begin{equation}
\widehat{\mathrm{EPFD}}_{0\,\mathrm{dBi}}(\mathbf{\hat{c}})
=
\sum_{i=1}^{N_{\mathrm{cells}}}
\bar{\Phi}_i\,
N\,\hat{B}_{\mathrm{synth}}(\mathbf{\hat{c}}\,|\,\mathbf{\hat{s}}_i)
\, B_{\mathrm{prim}}(\mathbf{\hat{s}}_i),
\label{eq:epfd0_meas_beam}
\end{equation}

which is the form implemented in the present analysis.  This results in an all-sky measurement-based EPFD map in which each output cell represents the aggregate interference level for that pointing direction, including contributions entering through both the main beam and sidelobes.



\begin{figure*}[t]
    \centering
    \includegraphics[width=\textwidth]{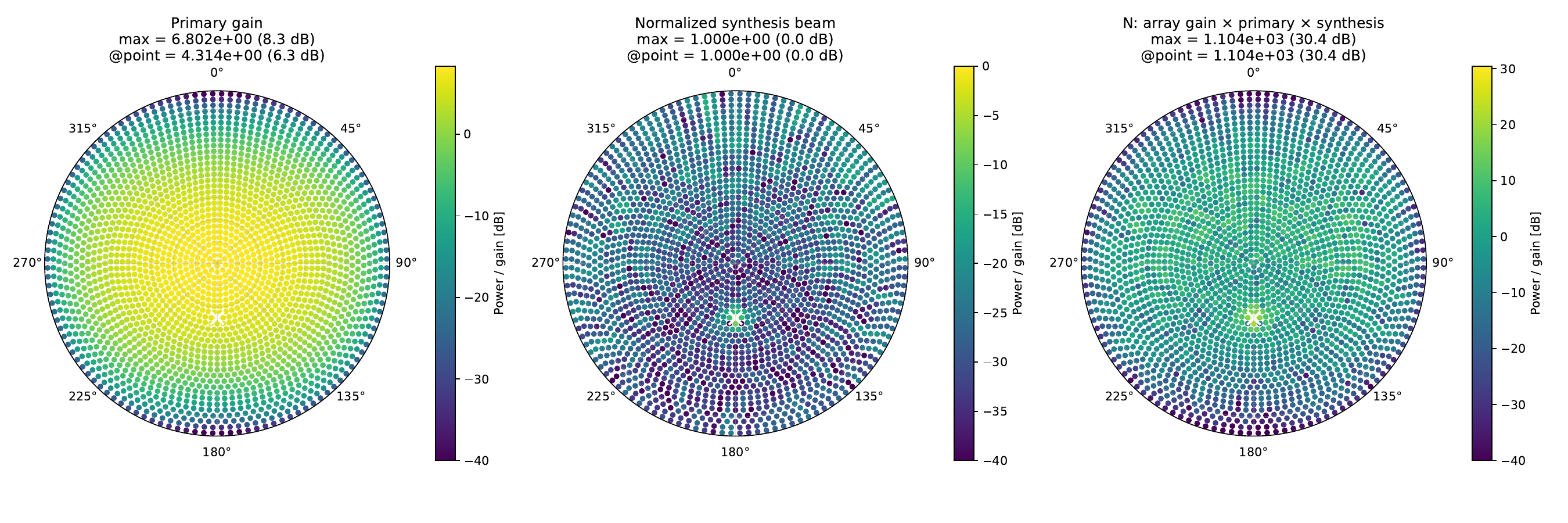}
    \caption{Primary beam, normalized synthesis beam, and combined receive response used in the measurement-based EPFD summation at 150~MHz. The combined response is evaluated as a function of source direction for a representative output pointing. The cross indicates the output pointing direction at Az = 180$^\circ$ and El = 61.5$^\circ$. The plot titles indicate the gain at the pointing direction as well as the maximum gain over the visible hemisphere. The beam responses are shown over the full visible hemisphere, although the nominal analysis only includes detections above 20$^\circ$ elevation, and the restricted analysis considers output sky cells above 40$^\circ$ elevation.}
    \label{fig:beam_components}
\end{figure*}

An example of the primary beam, normalized synthesis beam, and combined receive response is shown in Figure~\ref{fig:beam_components} for a representative output pointing at Az = 180$^\circ$ and El = 61.5$^\circ$. The combined response is evaluated as a function of source direction and defines the directional weighting applied to detected sources in the measurement-based EPFD summation.

The final dataset consists of one measurement-based EPFD sample per output sky cell per time window. Because the receiving response includes sidelobes, each active time window yields a fully populated all-sky EPFD map, provided that the window contains at least one detection. Each output sky cell therefore represents the aggregate interference level that would be obtained for that pointing direction using the detected source population. Empirical cumulative distribution functions (CDFs) are then constructed from these samples and compared to the RA.769 threshold.



\subsection{Results}
\label{subsec:results}



\begin{figure}
    \centering
    \includegraphics[width=\linewidth]{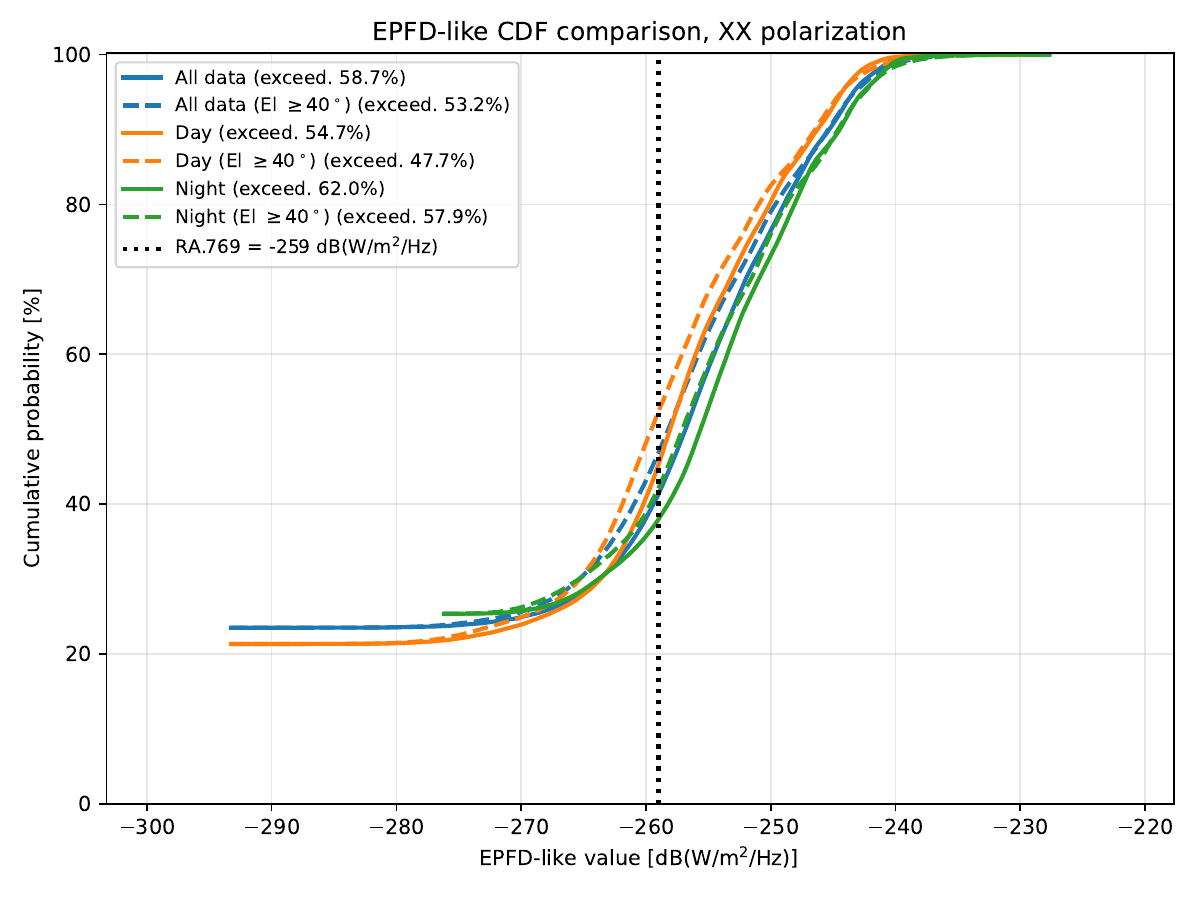}
    \caption{
    Measurement-based EPFD cumulative distributions for the XX polarization. Solid curves use all output sky cells, while dashed curves restrict the output EPFD cells to elevations above 40$^\circ$. The vertical dotted line indicates the RA.769 threshold of $-259$~dB(W\,m$^{-2}$\,Hz$^{-1}$).
    }
    \label{fig:cdf_xx}
\end{figure}

\begin{figure}
    \centering
    \includegraphics[width=\linewidth]{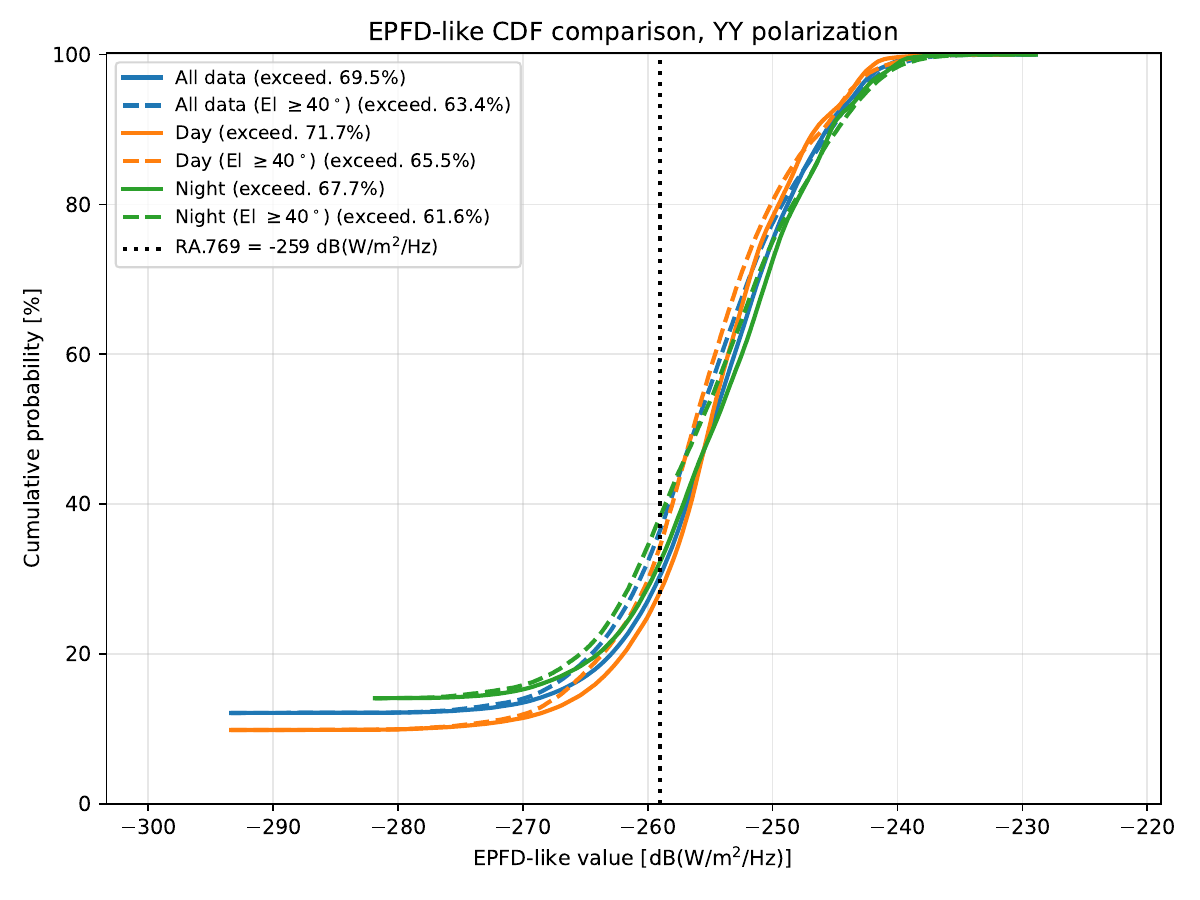}
    \caption{
    Same as Figure~\ref{fig:cdf_xx}, but for the YY polarization.
    }
    \label{fig:cdf_yy}
\end{figure}

The methodology described in Section~\ref{subsec:meth} was applied to the EDA2 detection dataset using observing channels overlapping the 150.05--153~MHz protected band, rather than to a strictly band-limited subset fully contained within that frequency range. After filtering, 12\,661 detections were retained for the XX polarization and 16\,853 detections for the YY polarization. 


The sky was discretized into 2\,292 cells following the M.1583 tessellation, corresponding to a typical solid angle of approximately $2.7\times10^{-3}$~sr ($\sim$9~deg$^2$) per cell.

Inspection of the azimuth/elevation distribution of detections (Figure~\ref{fig:azel_flux_distribution}) shows that the majority of detected signals are concentrated at elevations above approximately 40$^\circ$, where the sensitivity of the array is highest. Motivated by this behaviour, an additional analysis was performed using only output sky cells above 40$^\circ$ elevation, while retaining the nominal 20$^\circ$ elevation cut on detections, in order to evaluate the dependence of the measurement-based EPFD distributions on the selected output angular region.

The data span both day and night conditions. Across the three observing periods considered here, we identify 132 non-overlapping 2\,000~s windows in observing channels overlapping the 150.05--153~MHz protected band. Of these, 101 windows contain at least one XX detection and 116 contain at least one YY detection. We note that the fraction of windows not containing detections does not reflect the absence of satellites, but rather the observational and selection constraints of the dataset. In particular, the analysis is restricted to detected emissions above the instrument sensitivity and within a limited elevation range, reducing the number of windows containing observable signals.
Expanding all 132 windows across all sky cells yields 302\,544 window/cell samples for each polarization.
The windows were constructed without temporal overlap, such that each 2\,000~s segment represents an independent realization of the interference environment.
Following the convolution step, windows containing detections produce dense all-sky maps, while windows without detections remain zero-valued. For the nominal analysis, this yields fractions of non-zero samples of 76.5\% for XX and 87.9\% for YY in the full dataset.
Windows without detections are retained in the analysis as zero-valued sky maps, so the resulting distributions include both active and quiescent intervals within the observed frequency band.

The primary beam normalization, derived from a solid-angle integral over the sky, yields a peak gain of approximately 8.3~dBi, with a median gain close to 1~dBi and a minimum gain below $-50$~dBi at an elevation of 0$^\circ$. These values characterize the broad angular response of the individual array elements and are consistent across both polarizations.

The directional structure of the interferometric response, which includes both the main lobe and sidelobes, is governed by the synthesized beam of the array. Although the majority of the beam power is concentrated within the main lobe, a non-negligible fraction is distributed across the sidelobe structure. Contributions entering through sidelobes are therefore significantly attenuated relative to the main beam, but their large solid-angle coverage implies that they can collectively contribute to the aggregate interference level. This interplay between attenuation and sky coverage is consistent with the behaviour observed in the present dataset.

The resulting measurement-based EPFD cumulative distributions are shown in Figures~\ref{fig:cdf_xx} and \ref{fig:cdf_yy}. For the nominal analysis using all output sky cells, the XX polarization exceeds the RA.769 threshold in 58.7\% of samples for the full dataset, with day and night subsets yielding exceedance fractions of 54.7\% and 62.0\%, respectively. The corresponding YY exceedance fractions are 69.5\%, 71.7\%, and 67.7\%. These exceedance fractions are far above the 2\% single-system criterion in Recommendation ITU-R RA.1513-2.

Restricting the output EPFD cells to elevations above 40$^\circ$ reduces the exceedance fractions but does not change the qualitative conclusion. In this restricted analysis, the XX polarization gives exceedance fractions of 53.2\%, 47.7\%, and 57.9\% for the full, day, and night datasets, while YY gives 63.4\%, 65.5\%, and 61.6\%, respectively. This indicates that the exceedance is not driven solely by low-elevation output cells.

The same results can be expressed as \rev{conventional} margins to the 2\% criterion using Eq.~\ref{eq:margin_2pct}. For the nominal all-output-cell distributions, \del{the required uniform attenuation ranges from 16.3 to 18.3~dB.} \rev{the margins range from $-18.3$ to $-16.3$~dB.} When the output cells are restricted to elevations above 40$^\circ$, the corresponding margins range from \del{17.0 to 18.6~dB.} \rev{$-18.6$ to $-17.0$~dB.} \del{These margins provide a more direct compatibility interpretation; they quantify how far} \rev{Equivalently, a uniform attenuation of 16.3--18.6~dB would be required to bring} the 98th percentile of the measurement-based EPFD distribution \del{lies above} \rev{to} the RA.769 threshold.

\begin{table*}[t]
\centering
\label{tab:epfd_margins}
\begin{tabular}{llccc}
\hline
\textbf{Pol.} &
\textbf{Subset} &
\textbf{Output cells} &
\textbf{Exceedance} &
\del{\textbf{$M_{2\%}$ [dB]}} \rev{\textbf{$\Delta_{2\%}$ [dB]}} \\
\hline
XX & All data & All elevations & 58.7\% & \rev{-}17.6 \\
XX & Day & All elevations & 54.7\% & \rev{-}16.3 \\
XX & Night & All elevations & 62.0\% & \rev{-}18.2 \\
YY & All data & All elevations & 69.5\% & \rev{-}17.5 \\
YY & Day & All elevations & 71.7\% & \rev{-}16.6 \\
YY & Night & All elevations & 67.7\% & \rev{-}18.3 \\
\hline
XX & All data & El $\geq 40^\circ$ & 53.2\% & \rev{-}18.0 \\
XX & Day & El $\geq 40^\circ$ & 47.7\% & \rev{-}17.0 \\
XX & Night & El $\geq 40^\circ$ & 57.9\% & \rev{-}18.5 \\
YY & All data & El $\geq 40^\circ$ & 63.4\% & \rev{-}18.2 \\
YY & Day & El $\geq 40^\circ$ & 65.5\% & \rev{-}17.3 \\
YY & Night & El $\geq 40^\circ$ & 61.6\% & \rev{-}18.6 \\
\hline
\end{tabular}
\caption{
Measurement-based EPFD exceedance fractions and margins relative to the 2\% single-system criterion. The margin \del{$M_{2\%}$} \rev{$\Delta_{2\%}$} is computed as the \del{98th percentile of the EPFD distribution minus the} RA.769 threshold of $-259$~dB(W\,m$^{-2}$\,Hz$^{-1}$)\del{.} \rev{minus the 98th percentile of the EPFD distribution. Negative values therefore indicate that attenuation is required.}
}
\end{table*}

The distributions span a wide dynamic range and lie well above the RA.769 threshold over a significant fraction of samples. The elevation-restricted analysis produces slightly lower exceedance fractions than the nominal analysis, but slightly \del{larger 2\% compatibility margins}\rev{more negative compatibility margins, corresponding to slightly larger required uniform attenuation values.}. This indicates that restricting the output cells to higher elevations changes the shape of the upper tail without altering the qualitative conclusion. While the overall shapes of the XX and YY distributions are broadly similar, the YY dataset contains a larger number of detections (16\,853 compared to 12\,661 for XX),
resulting in a higher number of windows with detections (116 for YY and 101 for XX), and a corresponding increase in the fraction of non-zero samples. While the overall shapes of the XX and YY distributions remain broadly similar, the larger dataset reveals slightly stronger variations between subsets, particularly in the day/night comparison for the XX polarization.

The day and night subsets do not reveal a strong systematic dependence on solar illumination at the level of the full distribution. Moderate variations are observed, particularly for the XX polarization, where the exceedance fraction is higher during night-time conditions. The YY polarization is more stable between day and night. Overall, the high-percentile behavior remains broadly consistent across configurations, indicating that the upper tail of the distribution is dominated by strong emitters rather than purely diurnal effects.

\subsection{Comparison to previous simulations of low frequency UEMR}
\label{LOFAR_comp}

\cite{2023A&A...676A..75D} and \cite{2024A&A...689L..10B} previously reported the detection of UEMR from 68 Starlink satellites using the LOFAR telescope between 110 and 188 MHz, at spectral power flux densities of 0.1 to 10 Jy for broad-band radiation components, and from 10 to 500 Jy for narrow-band ($<$12.2 kHz) radiation components.  

In addition to reporting these measurements, the authors also performed indicative EPFD simulations, according to the ITU-R calculation description.  It is worth comparing these EPFD simulations to the results reported in this paper, which are based on a measurement-based EPFD estimate.  

In \cite{2023A&A...676A..75D}, the EPFD simulation proceeded by assuming a constellation of 4408 Starlink satellites and considered UEMR in the 150.05 - 153 MHz band allocated to RAS. Recognizing that the radiated power from an individual satellite is unknown for UEMR, a value for radiated power was assumed to be 30 dB[$\mu Vm^{-1}$], a typical radiation level found in the commercial standards, for example, CISPR-32 based on a 120 kHz bandwidth and a distance of 10 m from the source of radiation.  Further, this radiation was assumed to be isotropic.

The EPFD simulations also recognized that the representation of LOFAR as a low frequency array was a complex undertaking in the ITU-R framework, and thus a simplified receiving antenna was assumed, being 25 m and 70 m diameter dishes (corresponding approximately to an SKA-Low station and a LOFAR station, respectively).

Results for the Starlink EPFD simulation, based on these assumptions, appear in Figure 2 of \cite{2023A&A...676A..75D}, showing that the simulated Starlink constellation exceeded the ITU-R RA.1513-2 in 100\% of the simulated EPFD trials\rev{.The reported value of $-173.9\pm0.1$~dB(W\,m$^{-2}$) is expressed over the full 150.05--153~MHz RAS band, whereas the present work reports spectral power flux densities in dB(W\,m$^{-2}$\,Hz$^{-1}$). Using a bandwidth of 2.95~MHz, this corresponds to approximately $-238.6$~dB(W\,m$^{-2}$\,Hz$^{-1}$), assuming uniform scaling across the band. Equivalently, the value lies about 20~dB above the integrated RA.769 threshold of approximately $-194$~dB(W\,m$^{-2}$), or about 20~dB above the corresponding per-Hz threshold used here. This is broadly consistent with the present measurement-based results, for which the 98th percentile lies 16.3--18.6~dB above the RA.769 threshold. The comparison should nevertheless be interpreted cautiously because the earlier work used a forward simulation with assumed isotropic UEMR levels and simplified receiving antennas, whereas the present work uses measured source flux densities and reports both exceedance fractions and 98th-percentile margins.} \del{, with a peak power in the CDF of $-173.9\pm0.1$ dB[$Wm^{-2}$] (note, not the same units as used in this paper, which is the per Hz equivalent). However, the comparison should be interpreted cautiously because the earlier simulation results were summarized using different distribution statistics, whereas the present work reports both threshold exceedance fractions and the 98th-percentile margin associated with the 2\% single-system criterion.}

While the underlying assumptions and input information are significantly different between the \cite{2023A&A...676A..75D} simulations and the measurement-based EPFD calculation in this paper, both results show a significant exceedance of the ITU-R threshold power \rev{by a similar order of magnitude}, and therefore a reasonable level of consistency\del{.} \rev{despite the different assumptions and methodologies.}




\section{A proposed adaptation of the EPFD calculation for measurement data and a suggested instrumentation approach}
\label{sec:prop}


The results presented in this work highlight both the potential and the limitations of a measurement-driven approach to EPFD. While the methodology described in Section~\ref{subsec:meth} provides a practical analogue to the ITU-R EPFD calculation, it also raises the broader question of how such an approach could be generalized, standardized, and ultimately used to support regulatory frameworks. In this section, we outline a possible path toward adapting the EPFD formalism to measurement data, and discuss the corresponding instrumentation and operational requirements.


All-sky interferometric measurements provide a particularly well-suited starting point for such an adaptation. At low radio frequencies, the large primary beams of simple antenna elements allow a substantial fraction of the visible sky to be observed simultaneously. In an interferometric image, each pixel can be interpreted as the output of a synthesized beam pointed in that direction, effectively providing a set of simultaneous beamformed measurements across the sky. This property aligns closely with the cell-based nature of the EPFD calculation, enabling a direct mapping between image pixels (or groups of pixels) and sky cells. As a result, all-sky imaging offers an efficient and comparatively unbiased means of sampling the directional interference environment, without requiring mechanical scanning or sequential beam steering.  In particular, measurement-based approaches offer a means to assess assumptions related to satellite emission levels, propagation effects, antenna responses, source detectability, and temporal variability using real observational data.



Low-frequency interferometric arrays are particularly well suited to this task because their wide instantaneous fields of view enable simultaneous sampling of large sky areas with high cadence. Extending comparable methodologies to higher frequencies would become increasingly challenging due to the reduced field of view and increased instrumental complexity, implying that direct measurement-based EPFD estimation may only be practical in selected observational regimes.

Equally important as the hardware is the associated data processing pipeline. The transformation from raw measurements to measurement-based EPFD products must be standardized and reproducible. This includes calibration of the instrument response, mapping of measurements on to a predefined sky tessellation, application of antenna gain corrections, aggregation into temporal windows, and construction of cumulative distributions. In addition, the pipeline should incorporate auxiliary information such as satellite ephemerides (which would be best provided by operators), enabling the association of detected signals with specific systems when possible. The output of the pipeline should consist of a reduced set of products directly comparable to the EPFD formalism, including per-cell time-averaged power flux densities, cumulative distribution functions, and threshold exceedance statistics, along with metadata describing sensitivity, coverage, and completeness.

From a regulatory perspective, such a measurement framework would not replace existing EPFD calculations, but rather complement them. EPFD is currently used as a predictive tool in the coordination and filing process, and remains, to a large extent, a theoretical construct. Measurement-based estimates provide an empirical counterpart, capturing the aggregate behavior of real systems, including effects that are difficult to model accurately, such as unintended radiation or deviations from nominal operating conditions. In practice, measurement-derived EPFD products could be used to validate forward models, inform coordination discussions, and document the interference environment at specific sites. Over time, and subject to sufficient standardization, they could also contribute to compliance verification or post-deployment assessment of satellite systems.

The operation of such monitoring systems raises additional considerations regarding governance and data stewardship. To ensure credibility and transparency, instruments should ideally be operated by independent scientific or technical institutions, with clearly defined and publicly documented processing pipelines. At the same time, collaboration with regulators and satellite operators is essential to ensure that the measurements are interpretable and relevant to regulatory needs. A hybrid model, in which observational facilities provide the data and processing, while regulatory bodies define the standards and interpretation frameworks, may offer a practical path forward.

Overall, the development of a measurement-driven EPFD framework represents both a technical and institutional challenge. The results presented in this work demonstrate that existing radio astronomy instruments can already provide meaningful constraints on the interference environment in certain frequency ranges. Extending this capability into a systematic and regulator-relevant framework will require coordinated efforts in instrumentation, data processing, and standardization, but offers a promising avenue toward bridging the gap between theoretical compatibility studies and real-world spectrum management.

\section{Discussion and conclusions}
\label{sec:disc}

This work highlights the need for radio astronomy to describe the impact of Unintended Electro-Magnetic Radiation (UEMR) within the framework, language, and terminology of the ITU-R. While the phenomenon of UEMR is now well established observationally, its interpretation remains largely disconnected from the regulatory constructs that govern spectrum sharing. UEMR is not explicitly distinguished as a separate category within the ITU-R framework, and therefore does not naturally enter compliance assessments based on EPFD. In practice, such \rev{signals} \del{emissions} fall under the broader category of \del{unintended} radiation, which is conceptually addressed in the Radio Regulations but remains difficult to characterize and enforce in operational settings.

A central result of this paper is that all-sky interferometric measurements can be interpreted in a manner that is compatible with the EPFD formalism. By recognising that each pixel in an interferometric image is equivalent to a beamformed response in a given direction, we show that measurements can be used to construct measurement-based EPFD estimates without relying on forward modelling assumptions. This is particularly important in the case of \del{unintended} radiation \rev{not associated with wanted emissions}, for which the transmit antenna characteristics and emission mechanisms are generally not known, and therefore cannot be reliably incorporated into forward-model EPFD calculations.

We provide a practical pathway toward verification of EPFD calculations using real data. More broadly, we demonstrate that EPFD, while fundamentally a theoretical construct, can be meaningfully adapted to observational datasets, provided that the limitations of measurements are properly accounted for.

The need for such measurement-based approaches is reinforced by the increasing complexity of the regulatory landscape. The recent United States Federal Communication Commission Report and Order on modernizing spectrum sharing for satellite broadband proposes a shift away from traditional EPFD limits toward performance-based criteria and coordination-based mechanisms \citep{FCC2026_EPFD}. While this approach is motivated by demonstrated gains in spectral efficiency and system capacity, it effectively reduces the central role of EPFD as a regulatory constraint, at least within the United States. \del{Unintended radiation} \rev{Radiation not directly associated with intentional transmissions} is not explicitly addressed in this framework, and therefore does not enter into the performance-based compatibility assessments that replace EPFD limits. From a radio astronomy perspective, this creates a potential blind spot, as UEMR is largely isotropic and not associated with the highly directional beams that underpin the performance-based arguments for relaxing EPFD constraints. As a result, increases in the number of satellites, transmit powers, or constellations may translate more directly into increased levels of UEMR, without being captured by existing regulatory metrics.

In this context, our results provide an initial empirical baseline for UEMR levels in a protected radio astronomy band. The measurement-based EPFD distributions exceed the RA.769 threshold in approximately 5\del{9}\rev{0}--70\% of samples \del{in the nominal analysis}, depending on polarization\rev{,} \del{and} observing subset\rev{, and output-cell elevation selection. In the nominal analysis using all output sky cells, the exceedance fractions are approximately 55--72\%}. When restricting output cells to elevations above 40$^\circ$, the exceedance fractions remain high, at approximately 48--66\%. These values are far above the 2\% single-system criterion in Recommendation ITU-R RA.1513.

Expressed in the form commonly used in compatibility studies, the corresponding 2\% compatibility margins \rev{are negative, ranging from $-18.6$ to $-16.3$~dB. Equivalently, under the assumption of a uniform reduction applied to all measured source flux densities, an attenuation of 16.3--18.6~dB would be required to bring the corresponding distributions to the 2\% single-system criterion.} \del{indicate that the 98th percentile of the measurement-based EPFD distributions lies 16.3--18.6~dB above the RA.769 threshold. Equivalently, under the assumption of a uniform reduction applied to all measured source flux densities, an attenuation of 16.3--18.6~dB would be required to bring the corresponding distributions to the 2\% single-system criterion.}

Because the present estimator is affected by detection incompleteness, these margins should be interpreted as lower-bound estimates for the detected-source population. The rapid expansion of mega-constellations, both in terms of the number of satellites and the number of independent systems, suggests that aggregate effects will become increasingly important. A simple extrapolation indicates that an order-of-magnitude increase in the number of transmitting satellites is expected to lead to a corresponding shift in the upper tail of the EPFD distribution, potentially bringing it even further from regulatory thresholds. While such extrapolations are necessarily approximate, they highlight the importance of considering aggregate effects explicitly, particularly in the absence of formal regulatory limits on UEMR.

These considerations point to the need for regular and systematic monitoring of UEMR at frequencies protected for radio astronomy. Measurement-based EPFD analyses provide a natural framework for such monitoring, enabling consistent tracking of interference levels over time and across different observatories. Importantly, such monitoring should be designed to provide actionable information, both for satellite operators and spectrum regulators. For operators, measurements can identify specific emission characteristics, frequencies, or operational modes associated with elevated UEMR, supporting mitigation through design improvements or operational adjustments. For regulators, long-term datasets can inform evidence-based policy decisions, particularly in a context where existing regulatory frameworks are evolving. This need is consistent with existing provisions in the Radio Regulations, including Resolution 739 (Rev. WRC-19), which recognizes that unwanted emissions may arise from physical mechanisms that are not fully predictable prior to deployment.

The development of dedicated instrumentation and data processing pipelines is therefore a key component of this effort. All-sky imaging arrays operating in protected bands offer a particularly efficient approach, as they provide continuous, unbiased coverage of the visible sky, and naturally support the cell-based methodology underlying EPFD calculations. When combined with automated detection, satellite identification, and statistical analysis, such systems can deliver near real-time assessments of interference levels. Given the global nature of satellite constellations, a coordinated network of such instruments, operated by the radio astronomy community in collaboration with regulatory agencies, would provide the most comprehensive picture of UEMR impacts.

Ultimately, the management of UEMR will require continued collaboration between astronomers, satellite operators, and regulators. Measurement-based approaches can play a central role in this process by providing a shared, quantitative basis for discussion. The ability to report observed emissions, track their evolution over time, and relate them to specific systems or design choices, will be essential for developing effective mitigation strategies. As satellite constellations continue to evolve, such feedback loops between observation, analysis, and engineering will be critical to ensuring that the scientific use of the radio spectrum remains compatible with the rapid growth of space-based communication systems. In this context, measurement-driven frameworks provide a critical pathway for bridging the gap between regulatory definitions and the observed interference environment.

\begin{acknowledgements}
  The authors thank the anonymous referee for constructive comments that
  improved the manuscript. \rev{This work was supported by the National Science Foundation grant no.2229428.}
\end{acknowledgements}


\bibliographystyle{aa}   
\bibliography{references}

%





\end{document}